\documentclass[aps,showpacs,twocolumn,floatfix,superscriptaddress,nofootinbib]{revtex4}
\usepackage[dvips]{graphicx}
\usepackage{amsmath}
\usepackage{epsfig}

\begin{document}

\title{Neutrino mixing in matter}

\author{S.~H.~Chiu}
\email{schiu@mail.cgu.edu.tw}
\affiliation{%
Physics Group, CGE, Chang Gung University, Kwei-Shan 333, Taiwan}
\author{T.~K.~Kuo}
\email{tkkuo@purdue.edu}
\affiliation{%
Department of Physics, Purdue University, West Lafayette, IN 47907, USA}
\author{Lu-Xin~Liu}
\email{Luxin.Liu@wits.ac.za}
\affiliation{%
Department of Physics, Purdue University, West Lafayette, IN 47907, USA}
\affiliation{%
National Institute for Theoretical Physics,
Department of Physics and Centre for Theoretical Physics,
University of the Witwatersrand,
Wits, 2050,
South Africa}

\pagestyle{plain}
\pagenumbering{arabic}

\begin{abstract}
\noindent
Three-neutrino mixing in matter is studied through a set of evolution equations
which are based on a rephasing invariant parametrization.
Making use of the known properties of measured neutrino parameters, 
analytic, approximate solutions are obtained.
Their accuracy is confirmed by comparison with
numerical integration of these equations.
The results, when expressed in the elements squared of the mixing matrix,
exhibit striking patterns as the matter density varies.

\end{abstract}

\maketitle


It is well-established that neutrino mixing is modified 
by the presence of matter \cite{MSW}.  
Their effect has been used in the analyses of solar neutrinos,
and is expected to impact those of the supernova neutrinos,
when and if they become available.
Closer to home, there is a plethora of long baseline experiments
either in operation or in the planning stage. 
For these studies, it is essential to include the matter effects
in order to understand neutrino mixing at the fundamental level.

In the literature, effort has been devoted to solving problems
along this line \cite{ref2}. 
However, the process involves 
the complication of the cubic eigenvalue problems, 
and the results are usually far from transparent for a clear
extraction of the physical implications.

In this work we derive a set of evolution equations for the neutrino
parameters, using a rephasing invariant parametrization which
was developed for three-flavor quark mixing.   
The same formalism can be used in the neutrino sector, as long as it is
used for lepton number conserving processes, such as in neutrino oscillation,
which will be studied here.  It will be shown that the coupled
equations have simple, analytic, solutions which, 
when compared to the complete numerical solutions,
are quite accurate.  


For the neutrino mixing (PMNS) matrix ($V$),
we adopt the parametrization introduced earlier \cite{Kuo:05}. 
Briefly, without loss of generality, one can demand det$V=+1$. 
There are then a set of rephasing invariants
\begin{equation}\label{eq:g}
\Gamma_{ijk}=V_{1i}V_{2j}V_{3k}=R_{ijk}-iJ,
\end{equation}
where their common imaginary part can be identified with
the Jarlskog invariant $J$ \cite{Jar:85}.  
Their real parts were defined as
\begin{equation}
(R_{123},R_{231},R_{312};R_{132},R_{213},R_{321})
=(x_{1},x_{2},x_{3};y_{1},y_{2},y_{3}).
\end{equation}
These variables are bounded by $\pm 1$: 
$-1 \leq (x_{i},y_{j}) \leq +1$,
with $y_{j} \leq x_{i}$ for any ($i,j$). They satisfy two constraints
\begin{eqnarray}
\mbox{det}V=(x_{1}+x_{2}+x_{3})-(y_{1}+y_{2}+y_{3})=1, \\
(x_{1}x_{2}+x_{2}x_{3}+x_{3}x_{1})-(y_{1}y_{2}+y_{2}y_{3}+y_{3}y_{1})=0.
\end{eqnarray}
In addition, it is found that 
\begin{equation}\label{eq:J}
J^{2}=x_{1}x_{2}x_{3}-y_{1}y_{2}y_{3}.
\end{equation}

The $(x,y)$ parameters are related to $|V_{ij}|^{2}$ by
\begin{equation}\label{eq:w}
 W = \left[|V_{ij}|^{2}\right]
   = \left(\begin{array}{ccc}
                    x_{1}-y_{1} & x_{2}-y_{2}   &  x_{3}-y_{3} \\
                     x_{3}-y_{2} & x_{1}-y_{3}  & x_{2}-y_{1} \\
                    x_{2}-y_{3}  &   x_{3}-y_{1}    & x_{1}-y_{2} \\
                    \end{array}\right).  
\end{equation}
One can readily obtain the parameters $(x,y)$ from $W$ by computing its cofactors,
which form the matrix $w$ with $w^{T}W=(\mbox{det}W)I$, and is given by
\begin{equation}\label{eq:co}
 w = \left(\begin{array}{ccc}
                    x_{1}+y_{1} & x_{2}+y_{2}   &  x_{3}+y_{3} \\
                     x_{3}+y_{2} & x_{1}+y_{3}  & x_{2}+y_{1} \\
                    x_{2}+y_{3}  &   x_{3}+y_{1}    & x_{1}+y_{2} \\
                    \end{array}\right).     
\end{equation}


Physical measurables can always be expressed in terms of $(x,y)$.  For instance,
the $\nu_{\mu} \rightarrow \nu_{e}$ transition probability is given by
\begin{eqnarray}
& & P(\nu_{\mu} \rightarrow \nu_{e})=-4[F^{\mu e}_{21} \sin^{2}(\frac{D_{2}-D_{1}}{4E/L}) \nonumber \\
&+& F^{\mu e}_{31} \sin^{2}(\frac{D_{3}-D_{1}}{4E/L})
+F^{\mu e}_{32} \sin^{2}(\frac{D_{3}-D_{2}}{4E/L})] \nonumber \\
  & + & 8J\sin(\frac{D_{2}-D_{1}}{4E/L})\sin(\frac{D_{3}-D_{1}}{4E/L})\sin(\frac{D_{3}-D_{2}}{4E/L}),
  \end{eqnarray}
where $D_{i}=$ neutrino mass squared, $L$ is the length of baseline, $E$ is the neutrino energy, and
\begin{eqnarray}
x_{2}x_{3}+x_{1}y_{2}-y_{1}y_{2}-y_{2}y_{3} & \equiv & F^{\mu e}_{21}, \nonumber \\
-x_{1}x_{3}-x_{2}x_{3}+x_{3}y_{1}+y_{2}y_{3} & \equiv & F^{\mu e}_{31}, \nonumber \\
x_{1}x_{3}+x_{2}y_{3}-y_{1}y_{3}-y_{2}y_{3} & \equiv & F^{\mu e}_{32}.
\end{eqnarray}


Experimentally, the PMNS matrix in vacuum is well-approximated by    
\begin{equation}\label{w0}
W_{0} = \left(\begin{array}{ccc}
  \frac{2(1-\epsilon^{2})}{3}-2\eta & \frac{1-\epsilon^{2}}{3}+2\eta    &  \epsilon^{2} \\
 \frac{1+2\epsilon^{2}-\xi}{6}+\beta+\eta & 
     \frac{2+\epsilon^{2}-2\xi}{6}-\beta-\eta & 
     \frac{1-\epsilon^{2}+\xi}{2} \\
 \frac{1+2\epsilon^{2}+\xi}{6}-\beta+\eta  & 
\frac{2+\epsilon^{2}+2\xi}{6}+\beta-\eta  & 
    \frac{1-\epsilon^{2}-\xi}{2} \\
                    \end{array}\right)
\end{equation}    
with $(\epsilon, \eta, \beta, \xi) \ll 1$.  
$W_{0}$ reduces to the tri-bimaximal \cite{tribi} 
matrix when $\epsilon=\eta=\beta=\xi=0$.
If we allow the parameters $(\epsilon, \eta, \beta, \xi)$ to take on arbitrary values,
the matrix above can be used as a general parametrization of the mixing matrix.
Also, it is related to the familiar ``standard parametrization" \cite{data} 
by $S^{2}_{13}=\epsilon^{2}$, $S^{2}_{12}=\frac{1}{3}+\frac{2\eta}{1-\epsilon^{2}}$, 
$S^{2}_{23}=\frac{1}{2}+\frac{1}{2}\frac{\xi}{1-\epsilon^{2}}$,
and $\beta$ is a complicated function but $\beta \simeq \frac{\sqrt{2}}{3}C_{\phi}S_{13}$
for $(\epsilon, \eta, \xi) \ll1$.
From $W_{0}$, we find readily $x_{10}\simeq 1/3$, $x_{20} \simeq 1/6$, $x_{30} \simeq 0$,
and $x_{i0}+y_{i0}\simeq 0$ $(i=1,2,3)$.

In the flavor basis, the effective Hamiltonian for neutrinos
propagating in matter is given by $H_{\mbox{eff}}=\frac{H}{2E}$,
\begin{equation}\label{H}
H=\left[ V_{0}
                    \left(\begin{array}{ccc}
                    m_{1}^{2} &     &   \\
                              & m_{2}^{2} &  \\
                              &       & m_{3}^{2}  \\
                    \end{array}\right)
                    V_{0}^{\dag} + 
                         \left(\begin{array}{ccc}
                    A &     &   \\
                              & 0 &  \\
                              &       & 0  \\
                    \end{array}\right)\right],
\end{equation}
where $m_{1}$, $m_{2}$, and $m_{3}$ 
are the neutrino masses in vacuum, $V_{0}$ is the mixing matrix in vacuum,
and the induced mass $A=\sqrt{2}G_{F}n_{e}E$.

The matrix $H$ can be diagonalized, 
\begin{equation}
H=VDV^{\dag}=V \left(\begin{array}{ccc}
                    D_{1} &     &   \\
                              &D_{2} &  \\
                              &       & D_{3}  \\
                    \end{array}\right) V^{\dag},
\end{equation}
where $D_{i}=M_{i}^{2}$ is the squared mass in matter.
To study how the elements of $V$ evolve in matter, one may start with
$dH/dA$, 
\begin{equation}\label{dH}
\frac{dH}{dA}=\frac{d}{dA}[VDV^{\dag}]= 
  \left(\begin{array}{ccc}
                    1 &     &   \\
                              & 0  &  \\
                              &       & 0  \\
                    \end{array}\right).
\end{equation}
Eq.~(\ref{dH}) is then sandwiched by $V^{\dag}$ and $V$,
\begin{equation}\label{eq:matrix} 
V^{\dag}\frac{dV}{dA}D+ \frac{dD}{dA} +D \frac{dV^{\dag}}{dA}V=\left(\begin{array}{ccc}
                    |V_{11}|^{2} & V_{12}V_{11}^{*}   &  V_{13}V_{11}^{*} \\
                     V_{11}V_{12}^{*} & |V_{12}|^{2}  & V_{13}V_{12}^{*} \\
                    V_{11}V_{13}^{*}  &   V_{12}V_{13}^{*}    & |V_{13}|^{2} \\
                    \end{array}\right).
\end{equation}
Taking the diagonal and off-diagonal terms of Eq.~(\ref{eq:matrix}), 
and following the procedures in Ref.\cite{Chiu:09}, we find 
\begin{equation}\label{eq:di}
\frac{dD_{i}}{dA}=|V_{1i}|^{2}=x_{i}-y_{i}, \hspace{.2in} (i=1,2,3) 
\end{equation}
\begin{equation}\label{eq:dvi}
\frac{dV_{ij}}{dA}=\sum_{k\neq j} \frac{V_{ik}V_{1j}}{D_{j}-D_{k}} V_{1k}^{*}.
\end{equation} 
Eq.~(\ref{eq:dvi}) follows from $[(dV^{\dag}/dA)V]_{ik}=V_{1i}^{*}V_{ik}/(D_{i}-D_{k})$, $i \neq k$,
which can be inverted to solve for $dV/dA$ since the unknown element
$[(dV^{\dag}/dA)V]_{ii}$ is rephasing dependent, and can be set to vanish.


	\begin{table*}[ttt]
  \centering
	\begin{center}
 \begin{tabular}{cccc}  
 
 \hline

       & $1/(D_{1}-D_{2})$ & $1/(D_{2}-D_{3})$  & $1/(D_{3}-D_{1})$    \\ \hline
    $dx_{1}/dA$ & $x_{1}x_{2}-2x_{1}y_{2}+y_{1}y_{2}$  & 
    $-x_{1}x_{2}+x_{1}x_{3}+y_{1}y_{2}-y_{1}y_{3}$ & 
    $-x_{1}x_{3}+2x_{1}y_{3}-y_{1}y_{3}$          \\
     $dx_{2}/dA$  & $-x_{1}x_{2}+2x_{2}y_{1}-y_{1}y_{2}$  & 
     $x_{2}x_{3}-2x_{2}y_{3}+y_{2}y_{3}$  & 
     $x_{1}x_{2}-x_{2}x_{3}-y_{1}y_{2}+y_{2}y_{3}$     \\
   $dx_{3}/dA$ & $-x_{1}x_{3}+x_{2}x_{3}+y_{1}y_{3}-y_{2}y_{3}$ & 
   $-x_{2}x_{3}+2x_{3}y_{2}-y_{2}y_{3}$ & $x_{1}x_{3}-2x_{3}y_{1}+y_{1}y_{3}$     \\                     
   $dy_{1}/dA$ & $ -x_{1}x_{2}+2x_{2}y_{1}-y_{1}y_{2}$ &
   $-x_{1}x_{2}+x_{1}x_{3}+y_{1}y_{2}-y_{1}y_{3}$ & 
   $x_{1}x_{3}-2x_{3}y_{1}+y_{1}y_{3}$    \\
   $dy_{2}/dA$ & $x_{1}x_{2}-2x_{1}y_{2}+y_{1}y_{2}$  &
   $-x_{2}x_{3}+2x_{3}y_{2}-y_{2}y_{3}$  & 
   $x_{1}x_{2}-x_{2}x_{3}-y_{1}y_{2}+y_{2}y_{3}$ \\
   $dy_{3}/dA$ &  $-x_{1}x_{3}+x_{2}x_{3}+y_{1}y_{3}-y_{2}y_{3}$  & 
   $x_{2}x_{3}-2x_{2}y_{3}+y_{2}y_{3}$ & $-x_{1}x_{3}+2x_{1}y_{3}-y_{1}y_{3}$ \\
    $d(\ln J)/dA$  &  $-x_{1}+x_{2}+y_{1}-y_{2}$  &    $-x_{2}+x_{3}+y_{2}-y_{3}$
        & $x_{1}-x_{3}-y_{1}+y_{3}$              \\
   \hline
  \end{tabular}
    \caption{$dx_{i}/dA$, $dy_{i}/dA$, and $d(\ln J)/dA$ are expressed as sums
    of terms in $1/(D_{1}-D_{2})$, $1/(D_{2}-D_{3})$, and $1/(D_{3}-D_{1})$.}
  \end{center}
 \end{table*}


While Eq.~(\ref{eq:dvi}) is rephasing dependent, it may be used to compute
rephasing invariant quantities, $e.g.$, 
\begin{equation}
\frac{d\Gamma_{123}}{dA} =\frac{d}{dA}(V_{11}V_{22}V_{33})= \frac{dx_{1}}{dA}-i\frac{dJ}{dA}.         
 \end{equation}
After some algebra, separating the real and imaginary parts,
in addition to using different $\Gamma^{'}_{ijk}s$, 
we obtain the evolution equations for all $(x_{i},y_{i})$ and
$d\ln J/dA$, which are collected in Table I.
The evolution equations obtained here are entirely analogous to the familiar
RGE of mass matrices.  In both cases, the effective Hamiltonian contains a parameter, 
the energy for RGE, and $A$ for neutrino propagation. 
The respective evolution equations can be used to solve for eigenvalues and 
mixings as functions either of the energy scale, or of $A$.

The symmetric form of these equations allows us to find readily the result:
\begin{equation}\label{JD}
\frac{d}{dA}\ln[J(D_{1}-D_{2})(D_{2}-D_{3})(D_{3}-D_{1})]=0,
\end{equation}
$i.e.$, the product $[J(D_{1}-D_{2})(D_{2}-D_{3})(D_{3}-D_{1})]$ is a constant as $A$ changes,
a well-known result derived with different methods \cite{HSN}.  

In addition, by writing down the evolution equations for $(d/dA)\ln(x_{i}-y_{i})$, 
from Table I, we find another ``matter invariant":
\begin{equation}
\frac{d}{dA}[\frac{J^{2}}{(x_{1}-y_{1})(x_{2}-y_{2})(x_{3}-y_{3})}]=0.
\end{equation}                 
Or, $[J^{2}/(|V_{11}|^{2}|V_{12}|^{2}|V_{13}|^{2})]=\mbox{constant}$.
When we use the ``standard parametrization", it is seen that 
$J^{2}/(|V_{11}|^{2}|V_{12}|^{2}|V_{13}|^{2})=S^{2}_{\phi}S^{2}_{23}C^{2}_{23}$,
$i.e.$, $S_{\phi}\sin2\theta_{23}$ is independent of $A$, 
a result obtained earlier \cite{Toshev}.

The evolution equations for $(x,y)$ also have a structure
akin to that of the fixed point of single variable equations.    
It can be verified that, if $x_{i}+y_{i}=0$ $(i=1,2,3)$, then 
\begin{equation}\label{sigma}
\frac{d}{dA}(x_{j}+y_{j})=0,  \hspace{.2in} j=(1,2,3).
\end{equation} 
This result is understandable since the conditions
$x_{i}+y_{i}=0$ are equivalent to $W_{2i}=W_{3i}$, which,
in turn, imply that the effective Hamiltonian $H$ has a 
$\mu-\tau$ exchange symmetry \cite{23-sym}.
This symmetry is clearly independent of $A$ in Eq.~(\ref{H}),
resulting in Eq.~(\ref{sigma}).  Note also that there are actually
only two independent constraints in $x_{i}+y_{i}=0$.
Given any two of them, say for $i=1,2$, we can use Eq.(4)
to derive $x_{3}+y_{3}=0$.  Thus, the set of evolution
equations has a ``fixed surface", points on the
surface defined by $x_{i}+y_{i}=0$ stay on it as $A$ varies.

While analytical solutions to the equations in Table 1 are not available, 
as we will see, given the known physical parameters, one can exploit certain
characteristic properties thereof to arrive at simple,
but fairly accurate, solutions to these equations.

Experimentally, it is known that 
$\delta_{0}=m^{2}_{2}-m^{2}_{1} \cong 7 \times 10^{-5} eV^{2}$, 
$\Delta_{0}=m^{2}_{3}-m^{2}_{2} \cong 3 \times 10^{-3} eV^{2}$, so that
$\delta_{0}/\Delta_{0} \ll 1$ (we assume the ``normal" ordering of neutrino masses.
The ``inverted" case can be similarly treated).
Note that these values are relevant to long baseline
experiments since 
$A=\sqrt{2}G_{F}n_{e}E \sim (7.6 \times 10^{-5} eV^{2})(E/GeV)(\rho/g cm^{-3})$.


Since $\delta_{0} \ll \Delta_{0}$, we expect that the three-flavor problem
can be approximated by a pair of well separated two-flavor problems \cite{KP:89}. 
Indeed, the structure of the differential equations in Table I shows that
the variables $(x_{i},y_{i})$ evolve slowly as a function of $A$ except for
two regions, where $D_{1} \approx D_{2}$ and $D_{2} \approx D_{3}$,
corresponding to the two resonance regions.  
More precisely, let us denote by $(A_{0},A_{l},A_{i},A_{h},A_{d})$ the values of $A$
in vacuum $(A_{0}=0)$, at the lower resonance $(A_{l}, [d(D_{1}-D_{2})/dA]_{A_{l}}=0)$,
intermediate range $(A_{i})$, higher resonance $(A_{h}, [d(D_{2}-D_{3})/dA]_{A_{h}}=0)$,
and for dense medium $(A_{d})$.  Rapid evolution for $(x_{i},y_{i})$ only occurs for
$A\approx A_{l}$ and $A\approx A_{h}$.

For $0\leq A <A_{i}$, we need only to keep the terms 
$\propto 1/(D_{1}-D_{2})$ in Table I.
It is seen that
\begin{equation}
\frac{d(x_{1}-y_{2})}{dA}= \frac{d(x_{2}-y_{1})}{dA}= \frac{d(x_{3}-y_{3})}{dA}=0.
\end{equation}
Given $W_{0}$ (Eq.~(\ref{w0})), with $x_{i0}+y_{i0} \cong 0$,
Eq.~(\ref{sigma}) yields $x_{i}+y_{i} \cong 0$.  Thus, 
we expect that for $0 \leq A <A_{i}$, while the individual variables 
$(x_{1},x_{2},y_{1},y_{2})$ are rapidly changing, $x_{3} \simeq y_{3} \simeq \mathcal{O}(\epsilon)$
stay small, as do the combinations $x_{1}+y_{1} \cong x_{2}+y_{2} \cong 0$,
and $x_{1}+x_{2} \simeq \mbox{constant}$, $y_{1}+y_{2} \simeq \mbox{constant}$.
The differential equations can then be approximated, with $\delta \equiv D_{2}-D_{1}$, by 
\begin{equation}\label{eq:lde}
\frac{dx_{1}}{dA} \cong \frac{-4x_{1}x_{2}}{\delta} \cong  -\frac{dx_{2}}{dA},  \hspace{.2in}
\frac{d\delta}{dA} \cong 2(x_{2}-x_{1}).
\end{equation}              
It follows that
\begin{eqnarray}
& & \frac{d}{dA}[x_{1}x_{2}\delta^{2}] = 0, \nonumber \\  
& & \frac{d}{dA}[(x_{1}-x_{2})\delta] = -2(x_{1}+x_{2})^{2}\equiv -b_{l},
\end{eqnarray}
where $b_{l} \cong  2(x_{10}+x_{20})^{2}$. So, in the lower resonance region, 
the explicit, approximate, solutions are
\begin{eqnarray}\label{low}
\delta & = & [2b_{l}A^{2}-4c_{l}A+\delta^{2}_{0}]^{1/2}, \nonumber \\
x_{1} & = & \frac{1}{2}[(x_{10}+x_{20})-(b_{l}A-c_{l})/\delta], \nonumber \\
x_{2} & = & \frac{1}{2}[(x_{10}+x_{20})+(b_{l}A-c_{l})/\delta],
\end{eqnarray}
with $c_{l}=\delta_{0}(x_{10}-x_{20})$.
Also, $x_{1}+y_{1} \cong x_{2}+y_{2} \cong 0$, $x_{3} \cong y_{3} \cong 0$.
From $W_{0}$, we have $b_{l}=2(x_{10}+x_{20})^{2} \cong 1/2$.
We see thus, as $A$ sweeps through the lower resonance region, $\delta$ goes through
a minimum, $x_{1}$ decreases and $x_{2}$ rises while keeping $x_{1}+x_{2} \simeq 1/2$.
After the resonance, for large $A$ $(A \gg \delta_{0})$, $\delta \simeq A$, 
$x_{1} \rightarrow 0$, and $x_{2} \rightarrow 1/2$.
  
\begin{figure*}[ttt]
\caption{Numerical (solid) and approximate (dot-dashed) solutions for 
(a) all $D_{3}(A)$, $D_{2}(A)$, and $D_{1}(A)$, and 
(b) the enlarged plot of $D_{2}(A)$ and $D_{1}(A)$ in $0 \leq A/\delta_{0} \leq 10$.} 
\centerline{\epsfig{file=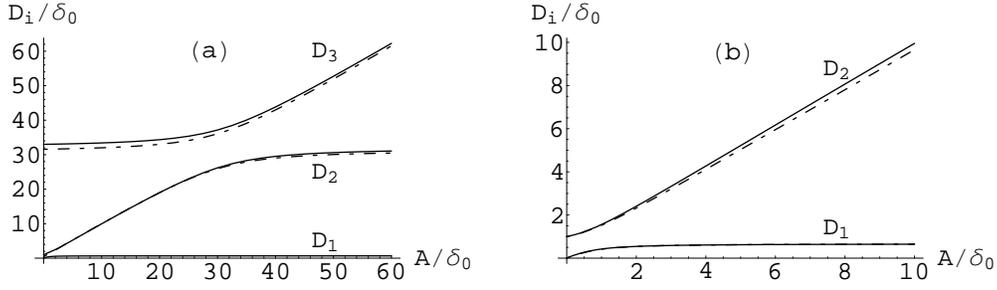,width=14 cm}}
\end{figure*} 

A similar analysis can be done for the region $A_{i} <A < A_{d}$.  Here, the starting values
($A_{l} \ll A <A_{h}$) are $x_{1} \simeq y_{1} \simeq 0$, $x_{3} \simeq y_{3} \simeq 0$,
$x_{2}\rightarrow 1/2$, $y_{2} \rightarrow -1/2$.
The differential equations are dominated by terms proportional to $1/(D_{2}-D_{3})$, and they satisfy
\begin{equation}
\frac{d(x_{1}-y_{1})}{dA}= \frac{d(x_{2}-y_{3})}{dA}= \frac{d(x_{3}-y_{2})}{dA}=0.
\end{equation}
With $\Delta \equiv D_{3}-D_{2}$, the approximate equations near $A \approx A_{h}$ are then
\begin{equation}\label{eq:hde}
\frac{dx_{2}}{dA} \cong \frac{-4x_{2}x_{3}}{\Delta} \cong  -\frac{dx_{3}}{dA},  \hspace{.2in}
\frac{d\Delta}{dA} \cong 2(x_{3}-x_{2}),
\end{equation}
together with $x_{2}+y_{2} \simeq x_{3}+y_{3} \simeq 0$,
while $x_{1}$ and $y_{1}$ are slowly varying so that $x_{1}\simeq y_{1} \simeq 0$ throughout.                
  
The solutions are
\begin{eqnarray}\label{high}
 & & \Delta = [2b_{h}A^{2}-4c_{h}A+\Delta_{0}^{2}]^{1/2}, \nonumber \\
 & & x_{2} = \frac{1}{2}[(x_{20}+x_{30})-(b_{h}A-c_{h})/\Delta], \nonumber \\
 & & x_{3} = \frac{1}{2}[(x_{20}+x_{30})+(b_{h}A-c_{h})/\Delta],
\end{eqnarray}
with $b_{h}=2(x_{20}+x_{30})^{2}$ and $c_{h}=\Delta_{0}(x_{20}-x_{30})$.

Thus, as $A$ goes from $A_{i}$ through $A_{h}$ to $A_{d}$, the changes for $(x_{i},y_{j})$ are:
$x_{2}\simeq 1/2 \rightarrow 0$; $y_{2}\simeq -1/2 \rightarrow 0$;
$x_{3}\simeq 0 \rightarrow 1/2$; and $y_{3} \simeq 0 \rightarrow -1/2$.

\begin{figure}[ttt]
\caption{The numerical (solid) and approximate (dot-dashed) solutions
for $x_{1}(A)$, $x_{2}(A)$, and $x_{3}(A)$.  Note that $y_{i}(A) \simeq -x_{i}(A)$.} 
\centerline{\epsfig{file=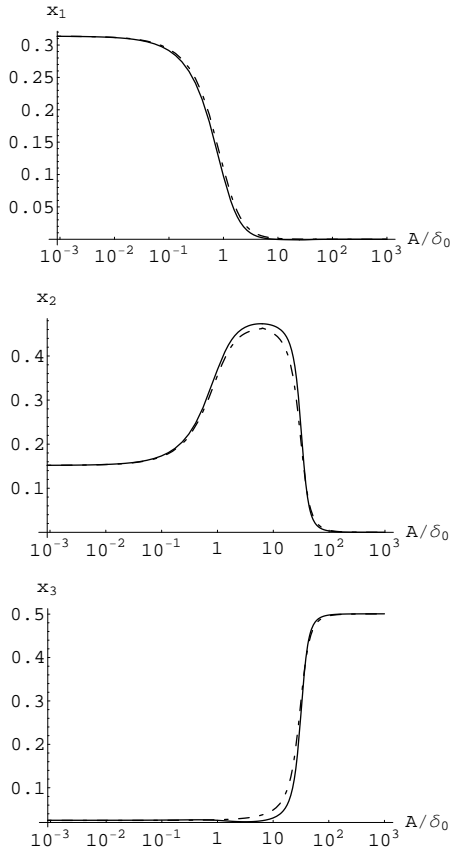,width= 6 cm}}
\end{figure} 

Our results can be summarized by the matrices $W$ at $A=(A_{0},A_{l},A_{i},A_{h},A_{d})$:
\begin{eqnarray}\label{eq:sum}
W_{0}  & \cong & \left(\begin{array}{ccc}
                    2/3 & 1/3   &  0 \\
                    1/6 & 1/3  & 1/2 \\
                   1/6  &   1/3   & 1/2 \\
                    \end{array}\right), \hspace{.15in}
    W_{l}  \cong  \left(\begin{array}{ccc}
                   1/2 & 1/2   &  0 \\
                     1/4 & 1/4  & 1/2 \\
                    1/4  &   1/4    & 1/2 \\
                    \end{array}\right),  \nonumber \\ 
 W_{i} & \cong & \left(\begin{array}{ccc}
                    0 & 1   &  0 \\
                     1/2 & 0  & 1/2 \\
                    1/2  &   0    & 1/2 \\
                    \end{array}\right), \hspace{.15in} 
   W_{h}  \cong  \left(\begin{array}{ccc}
                    0 & 1/2   &  1/2 \\
                     1/2 & 1/4  & 1/4 \\
                    1/2  &   1/4    & 1/4 \\
                    \end{array}\right),  \nonumber \\ 
   W_{d}  & \cong & \left(\begin{array}{ccc}
                    0 & 0   &  1 \\
                    1/2 & 1/2  & 0 \\
                    1/2  &  1/2   & 0 \\
                    \end{array}\right).
\end{eqnarray}
Together, these matrices exhibit the remarkable simplicity of the 
PMNS matrix as $A$ varies from $0$ to $\infty$.  
Note that all of the matrices have 
at least one zero, $W_{1I}=0$, implying $x_{I}=y_{I}=0$.
Also, they have
equal elements in their second and third rows,
$W_{2i}=W_{3i}$, so that 
$w_{1i}=0$ or $x_{i}+y_{i}=0$.  
As a consequence, using the unitarity conditions, the $W$ matrix
is completely fixed by its elements in the first row, $W_{1i}$.
These elements, in turn, control $dD_{i}/dA$, Eq.~(\ref{eq:di}).
Thus, the progression of $W$ as a function of $A$ can be read off from the plot 
of $D_{i}(A)$, which is given in Fig. 1.

It is straightforward to numerically integrate the evolution
equations for $(x,y)$.  To do this we choose the initial values
(in $W_{0}$) $\epsilon=0.17$, $\beta =0.02$,
corresponding to the experimental bounds
$|V_{e3}|^{2} \leq 0.03$ \cite{data} and an assumed CP violation
phase $\cos \phi=1/4$.  Also, $\xi=\eta=0$.  In Fig. 2 the results are compared to
the approximate solutions obtained earlier (Eqs.~(\ref{low}) and (\ref{high})).
The agreements are quite good.
The evolution of $J^{2}$ is shown in Fig. 3.  Compared to its vacuum value,
it is seen that, except for some enhancement near $A=A_{l}$,
$J^{2}$ tends to decrease with increasing $A$, as one would expect 
from Eq.~(\ref{JD}).

\begin{figure}[ttt]
\caption{The evolution of $J^{2}$ from the numerical (solid) and the approximate
(dot-dashed) solutions.} 
\centerline{\epsfig{file=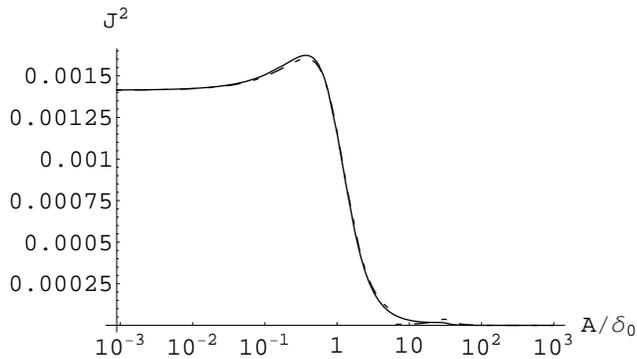,width=8.5 cm}}
\end{figure} 

 
In conclusion, in this work we derived the evolution equations for the 
neutrino parameters as a function of matter density.
We found  analytic, approximate, but simple solutions of these
equations for values centered around the known neutrino parameters.
This is possible because of two fortuitous circumstances: 
1) the neutrino mass differences are widely separated,
enabling one to use the two-flavor resonance approximation;
2) the mixing in vacuum satisfies $x_{i0}+y_{i0} \cong 0$,
which happens to lie on the ``fixed surface" of the evolution equations,
so that $x_{i}+y_{i} \cong 0$ for all $A$ values.
These solutions are summarized in Eq.~(\ref{eq:sum}), which exhibits
the extraordinary simplicity of $W$ as a function of $A$.  These results are 
found to be quite accurate when we compare them to those obtained by numerical
integration of the equations.
It is hoped that they will be useful in assessing the matter effects in
connection with the long baseline experiments, as well as efforts to 
determine CP-violation in the leptonic sector.                  
                   
S.H.C. is supported by the National 
Science Council of Taiwan, grant No. NSC 98-2112-M-182-001-MY2.

\end{document}